\documentclass[showpacs,preprintnumbers,amsmath,amssymb,aps,onecolumn,11 pt]{revtex4-1}

\usepackage{graphicx}
\usepackage{bm}
\usepackage{hyperref}
\usepackage{cases}
\usepackage{color}

\begin{document}


\title{{Coherent perfect absorption in deeply subwavelength films in the single photon regime.} }

\author{Thomas Roger$^{1}$}
\author{Stefano Vezzoli$^{2}$}
\author{Eliot Bolduc$^{1}$}
\author{Joao Valente$^{3}$}
\author{Julius J. F. Heitz$^{1}$}
\author{John Jeffers$^{4}$}
\author{Cesare Soci$^{2}$}
\author{Jonathan Leach$^{1}$}
\author{Christophe Couteau$^{2,5,6}$}
\author{Nikolay Zheludev$^{2,3}$}
\author{Daniele Faccio$^{1}$} \email{Corresponding d.faccio@hw.ac.uk}

\affiliation{$^{1}$ Institute for Photonics and Quantum Sciences and SUPA, Heriot-Watt University, Edinburgh, UK}
\affiliation{$^{2}$ Centre for Disruptive Photonic Technologies, Nanyang Technological University, Singapore}
\affiliation{$^{3}$ Optoelectronics Research Centre $\&$ Centre for Photonic Metamaterials, University of Southampton, UK}
\affiliation{$^{4}$ Department of Physics, University of Strathclyde, Glasgow, UK}
\affiliation{$^{5}$CINTRA CNRS-NTU-Thales, UMI 3288, Singapore}
\affiliation{$^{6}$Laboratory for Nanotechnology, Instrumentation and Optics, ICD CNRS UMR 6281, University of Technology of Troyes, Troyes, France}

\date{\today}

\begin{abstract}
{The technologies of heating, photovoltaics, water photocatalysis and artificial photosynthesis depend on the absorption of light and novel approaches such as coherent absorption from a standing wave promise total dissipation of energy. Extending the control of absorption down to very low light levels and eventually to the single photon regime is of great interest yet remains largely unexplored. Here we demonstrate the coherent absorption of single photons in a deeply sub-wavelength 50\% absorber. We show that while absorption of photons from a travelling wave is probabilistic, standing wave absorption can be observed  deterministically, with nearly unitary probability of coupling a photon into a mode of the material, e.g. a localised plasmon when this is a metamaterial excited at the plasmon resonance. These results bring a better understanding of the coherent absorption process, which is of central importance for light harvesting, detection, sensing and photonic data processing applications.}
 \end{abstract}

\maketitle

Recent studies provided unexpected but strong evidence that the quantum properties of light are conserved when photons are converted into surface plasmon polaritons, paving the way for active and ultrafast quantum plasmonic technologies \cite{Altewischer, Chang, Chang2, Akimov, Heeres, Tame, Fakonas} and stimulating a broad interest for the topic. At the same time, light interaction with nanostructured and nanotextured materials exploiting plasmonic resonances is a rapidly growing field of research with potential applications in photovoltaics \cite{Atwater}, water photocatalysis \cite{Linic, Pincella}, artificial photosynthesis \cite{Wang, Lee}, light harvesting for heating and light sources with engineered emissivity \cite{Liu}. Furthermore, plasmonics plays a crucial role in electromagnetic coupling of molecules to quantum dots and metal nanoparticles or nanowires \cite{Akimov, Kinkhabwala}. Different schemes have been developed to achieve strong coupling of light to metamaterials \cite{Fedetov, Schwanecke, Liu2, Thongrattanasiri} including the coherent perfect absorption that was first demonstrated in slabs of lossy materials \cite{Chong, Wan, Dutta-Gupta}. Light-with-light modulation based on the coherent perfect absorption in metamaterial films of sub-wavelength thickness is also possible and it has now been demonstrated with a continuous wave (CW) laser \cite{Zheludev} and with femtosecond optical pulses exhibiting modulation bandwidths of a few terahertz and possibly beyond \cite{Fang}. In this process, two coherent beams of light interact on a layer of plasmonic metamaterial in such a way that one beam modulates the intensity of the other. The interference of the two beams can eliminate the plasmonic Joule losses of light energy in the metamaterial with full transmission of the incident light. Depending on the mutual phase of incident beams it can also lead to total absorption of light. This provides a method to go beyond the theoretical limit of 50\% absorption in a thin film \cite{Zheludev}, but also a new way for controlling optical signals \cite{Mousavi}. In this work we explore the mechanisms of coherent absorption at the single photon level in deeply sub-wavelength films. We demonstrate that a single photon can be coupled to a plasmon mode of a metamaterial or absorbed in a multilayered graphene film  with nearly 100\% probability.\\

\section{Results}

\begin{figure}[ht]
\begin{centering}
\includegraphics[width = 0.8\textwidth]{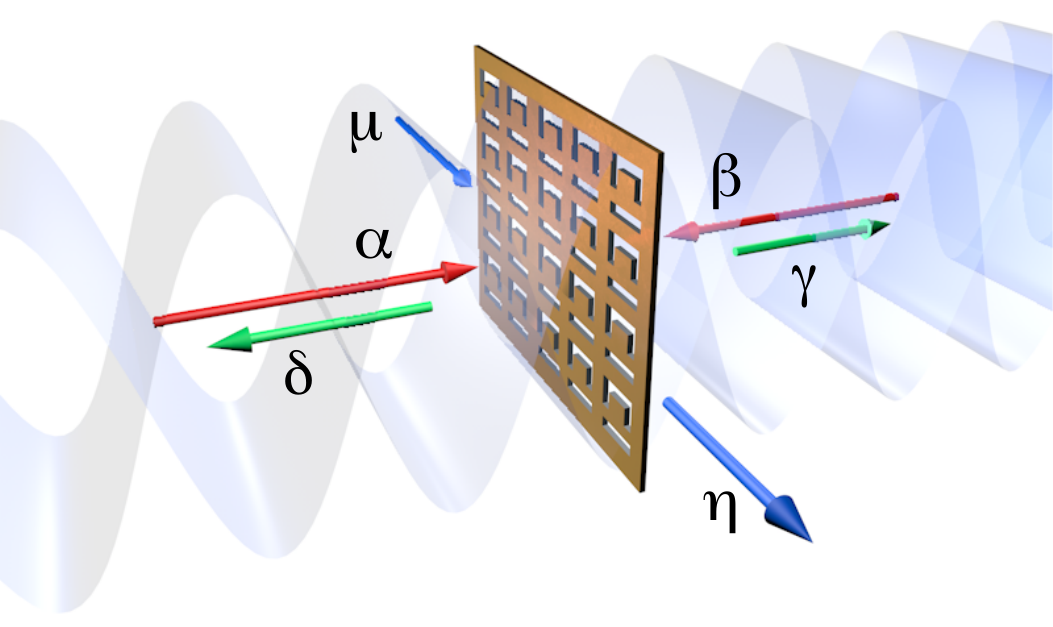}
\caption{\textbf{Schematic of metamaterial input/output ports.} Interaction of two coherent beams in a thin absorber, here represented as a plasmonic metamaterial: The film is described as a lossy beam-splitter with two input photon channels $\alpha$ and $\beta$, two photon output channels $\gamma$ and $\delta$ and plasmon input and output channels $\mu$ and $\eta$. }
\label{fig1}
\end{centering}
\end{figure}

\textbf{Coherent control in the single photon regime.} In a classical wave optics description, a thin absorbing film of sub-wavelength thickness experiences no interaction with light if it is placed in the node of a standing wave formed by two counter-propagating coherent waves of the same amplitude (e.g. obtained using a 50/50 beamsplitter).  Indeed, the electric field of the light has zero amplitude in the node making no contribution to dipole interactions in the medium. In contrast, the film absorbs strongly in the antinode of the standing wave where the magnitude of the oscillating field is at maximum. A film that absorbs 50\% for the traveling wave will absorb 100\% in the antinode of the standing wave \cite{Zheludev, Jeffers}. In the quantum regime, the {absorber} can be treated as a device with two input photon ports, $\alpha$ and $\beta$, and two output photon ports, $\gamma$ and $\delta$. We also assume that absorption is related to the excitation of a mode {in the material e.g. a plasmonic mode in the case of a metamaterial thus adding two more ports $\mu$ (input) and $\eta$ (output) to the description of the absorption process}, see Fig.~\ref{fig1}.\\

In the regime of linear optics the output photonic states $\gamma$, $\delta$ and plasmonic states $\eta$ are a linear combination of the input states $\alpha$, $\beta$ and $\mu$: \\

\begin{equation}
\left(\begin{array}{c}\gamma\\ \delta \\  \eta \end{array}\right) = S\left(\begin{array}{c} \alpha\\  \beta\\ \mu  \end{array}\right)
\end{equation}\\

\noindent where $S$ {is the scattering matrix that relates the input to the output states  and is given explicitly in the Methods}.\\

\begin{figure}[ht]
\includegraphics[width = 0.8\textwidth]{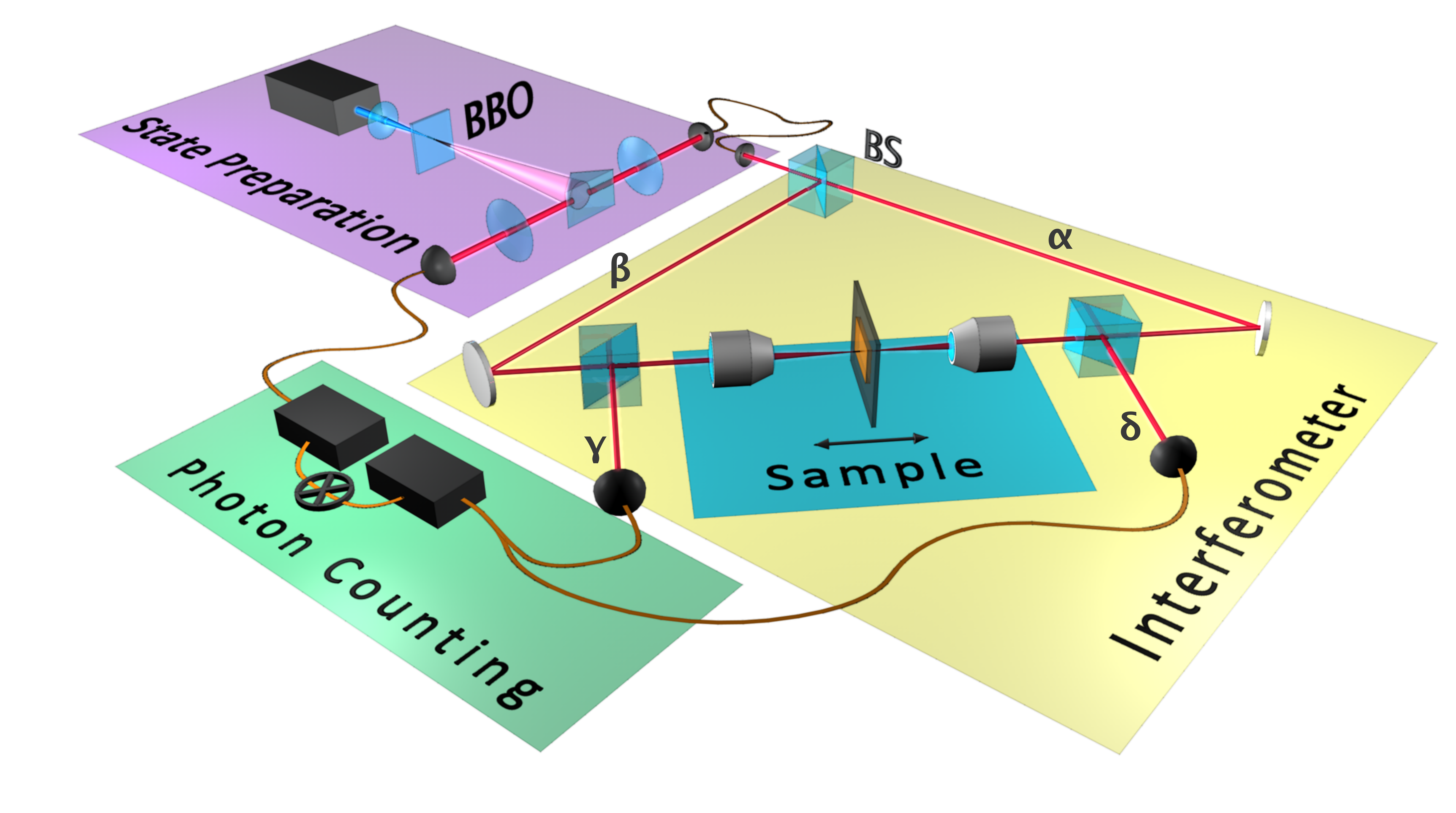}
\caption{\textbf{Single photon experiment on perfect coherent absorption}. Illumination of a type 1 beta-barium borate (BBO) crystal by a continuous wave $\lambda = $ 405 nm laser producing correlated single photon pairs by spontaneous parametric down conversion. The correlated photon pair are separated by a knife edge prism.  One photon of the correlated pair is used to herald the presence of the other photon that is launched into the interferometer. The metamaterial absorber is placed in the middle point of the interferometer and translated along the optic axis by a piezo-electrically actuated stage. The single photons are focussed onto the sample by 10x objectives. Photons are then detected in coincidence with the heralding photon at outputs $\delta$ and $\gamma$.}
\label{fig2}
\end{figure}

\textbf{Experiments with single photon Fock states.} In the experiment (layout shown in Fig.~\ref{fig2}),  a single photon launched into the interferometer via a lossless 50/50 beam-splitter generates a coherent superposition state at a metamaterial film. This state may be written as $|\psi\rangle = |\alpha\rangle + e^{i\phi}|\beta\rangle$, where $\phi$ is the phase shift between two input channels. {By tuning properly the reflectivity and the transmissivity parameters of the scattering process at the metamaterial sample, this state evolves} in such a way that the probability to observe at least one photon in either of the output channels $\gamma$ and $\delta$ is given by the expression, $P = |1/(2\sqrt{2})(1-e^{i\phi})|^2 = 1/4(1-\cos\phi)$. This implies that for $\phi = \pi$ no photons will be measured in the output channels and correspondingly the single input photon will be totally absorbed with 100\% probability. \\

\begin{figure}[h!]
\begin{centering}
\includegraphics[width = 16cm]{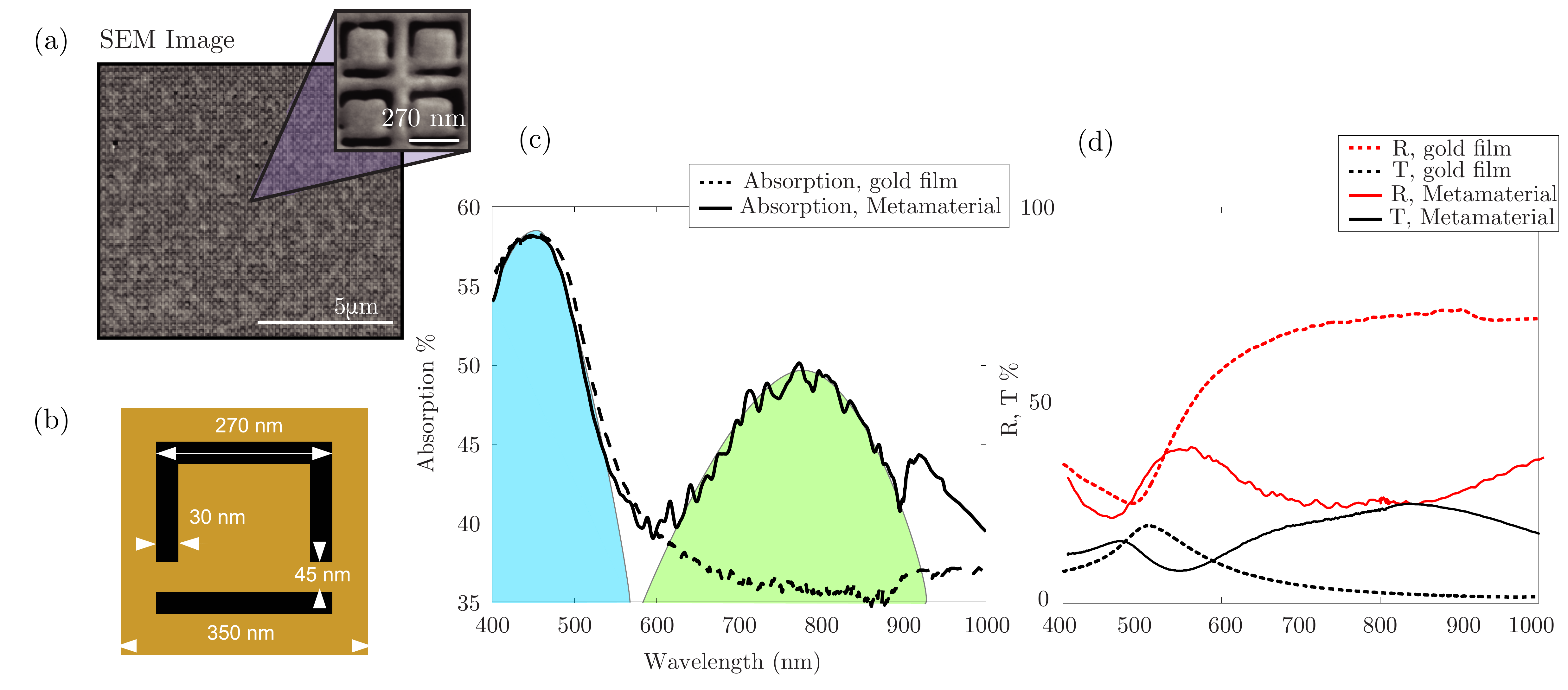}
\caption{Metamaterial absorber. (a) scanning electron microscope image of the milled 50 nm free-standing gold film; (b) Design of the unit cell; (c) The absorption spectrum of the metamaterial film (solid black line). The blue shaded region indicates the natural absorption resonance of gold around 450 nm whilst the green shaded region highlights the structure-induced resonance around 810 nm. For comparison, the absorption of an unstructured gold film of the same thickness is also shown (dashed black curve). (d) The reflection and transmission curves of the unstructured gold film (dashed curves) and of the metamaterial (solid curves): as can be seen, the structured material exhibits equal 25\% transmission and reflection around 800 nm, corresponding also to 50\% absorption.}
\label{figs1}
\end{centering}
\end{figure}

To demonstrate the effect of deterministic single photon-to-plasmon coupling and thus perfect absorption, we conducted the experiment with a plasmonic metamaterial exhibiting close to 50\% traveling wave absorption. The metamaterial absorber was manufactured by focused ion beam milling to form a free-standing 50nm thick gold film perforated with an array of asymmetric split ring structures, which provide the desired optical properties congruent with coherent perfect absorption, {i.e. 50\% absorption and equal reflection/transmission amplitudes} (see Fig. \ref{figs1} for a details of the sample design and optical characteristics).\\

The sample was placed at the centre point of the interferometer. Single photon states were prepared by spontaneous parametric down conversion (SPDC) of a laser diode with emission line cantered at the wavelength of 405 nm. We used a beta-barium borate ($\beta$-BBO) crystal producing non-collinear, degenerate entangled photon pairs at $\lambda = 810$ nm.  The output ports $\gamma$ and $\delta$ were monitored with single photon avalanche detectors gated by the heralding photon channel. The confidence with which a single photon in the  interferometer was heralded by a second photon were evaluated in a Hanbury-Brown-Twiss experiment that {returned a very high degree of heralded second-order coherence (see Methods)}, thus indicating a high fidelity of the single photon source. The relative phase shift between the input channels, $\phi$, was controlled by translating the metamaterial film along the light propagation direction with a piezoelectric actuator.\\

\begin{figure}[ht!]
\begin{centering}
\includegraphics[width = 0.8\textwidth]{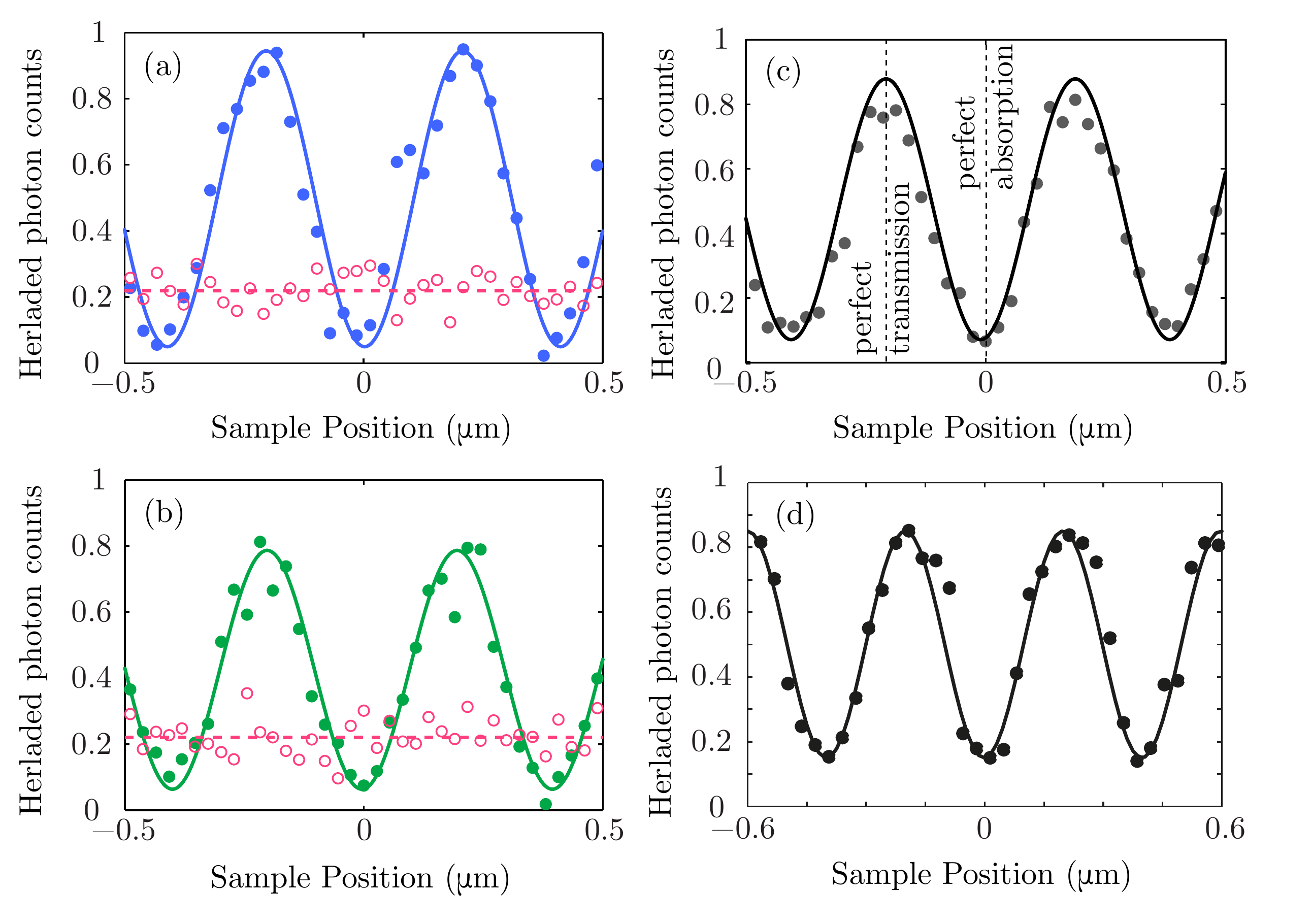}
\caption{Single photon perfect coherent absorption. (a) and (b) show the output photon count rates in channels $\gamma$ and $\delta$ normalised to the input photon count rates  in  channels $\alpha$ and $\beta$ as functions of the metamaterial absorber position along the standing wave (round, full symbols).  Also shown are the results of measurements when the input channel $\beta$ is blocked (open circles and dashed horizontal fitting lines).  (c) shows the half-sum of the normalized rates in channels $\gamma$ and $\delta$. The vertical dotted lines indicate the positions of nodes and anti-nodes, corresponding to almost perfect transmission and absorption regimes, respectively. (d) is the same as (c) but for a different 50\% absorber sample, made of a 30-layer chemical vapour deposition (CVD)-grown graphene film.}
\label{fig3}
\end{centering}
\end{figure}

Figures~\ref{fig3}(a) and (b) show the output photon count rate in channels $\gamma$ and $\delta$ normalised to the input photon count rate in channels $\alpha$ and $\beta$ {(measured on the same detectors by removing the sample)} as functions of the plasmonic absorber position. With a single photon entering the device at a time, we observe periodic oscillation in the output photon count rate as the phase shift $\phi$ between two input channels changes. The oscillation period of 405 nm corresponds exactly to the $\lambda/2$ period. Here perfect absorption corresponds to the minima of the curve. The overall modulation [Fig.~\ref{fig3}(c)] was measured to be between ~ 90\% and ~10\%: the shortfall from a 100\% modulation is explainable by a diffusive scattering from the metamaterial film fabrication imperfections and a contribution of quadrupole transitions to the absorption spectra of the split-ring metamaterial \cite{Khardikov}. Indeed, the efficiency of the quadrupole absorption is proportional to the gradient of the electric field of the standing wave that reaches maxima at its nodes, thus the presence of quadrupole absorption prevents a total collapse of dipole absorption at these positions. \\

\textbf{50 $\%$ absorbing metamaterial.} We stress that the optical response of the asymmetric split ring array is in general quite complex. Apart from the dominant dipole contribution of the optical response, there is also a dependence on the magnetic and electric quadrupole modes of the metamaterials resulting in the Fano-type absorption spectrum. This has been studied in a number of papers such as the one given in Ref. \cite{Khardikov}. De facto, the electric dipole response generates a different phase shift for transmitted and reflected waves in comparison with the magnetic and quadrupole responses. As a result of interferences of these responses, the pattern of absorption maxima and minima will be shifted with respect to the anti-nodes and nodes of the standing wave. However, the effect of perfect absorption will still be observed in much the same way as it would be observed for a metamaterial with solely dipolar response.\\


Finally, our set up can be converted for the study of absorption from a traveling wave by blocking the input channel $\beta$. In this case we register a position-independent level of  the normalised photon count rate  in the output channel $\gamma$, that indicates the probability of photon absorption [open circles in Fig.~\ref{fig3}(a)]. Indeed, by removing the Ôwhich-pathÕ ambiguity we are left with only the $\alpha$-state: the output signal $\gamma$ is simply determined by the traveling wave transmission coefficient of the metamaterial.  We obtain similar results by monitoring the output port $\delta$ [open circles in Fig.~\ref{fig3}(b)].\\

\section{Discussion}

We note that absorption in the free-standing metamaterial film has a predominantly plasmonic nature related to its nanostructuring. Therefore the coherent absorption process implies a nearly 100\% efficient coupling of the single photon to the plasmonic mode of the absorber. Hence, our experiment shows that although the absorption process of a single photon in a traveling wave is by its very nature probabilistic, absorption can be made completely deterministic by providing a Ôwhich-pathÕ ambiguity of the standing wave.\\
{It is important to note that the results shown here can be replicated with any material that exhibits the fundamental properties of 50\% absorption and deeply subwavelength thickness. To this end, we repeated the experiment with a 30-layer graphene film, which as can be seen in Fig.~\ref{fig3}(d) delivered very similar results.\\}
Our findings also expose the underlying quantum mechanism of optical gating via the coherent absorption process, which has previously been reported with CW and pulsed signals at classical light levels \cite{Zheludev, Fang}. The fact that modulation of light can be demonstrated with a single photon proves that the effect of modulation here does not rely upon one photon modulating another e.g. via a nonlinearity of the film. Rather, the coherent absorption gate exploits a difference in the absorption probabilities between the two configurations of the gate when the control beam is blocked or open. At higher photon fluxes (i.e. at classical light levels), this takes the form of an interference-controlled re-distribution of the energy flow between the inputs, outputs and the dissipative channel provided by the 50\% absorber \cite{Davis}. In contrast with a gate based on the materialÕs nonlinearity, the coherent absorption gate operates with no harmonic distortion and works at any intensity level.\\

In conclusion we have demonstrated experimentally that the coherent absorption process in a thin absorber holds at the single quantum level and that a single photon can be deterministically coupled to a plasmonic mode of a metamaterial. {More specifically, our results explicitly show that the same coherent absorption observed in the classical regime can also be observed with single photons. This  paves the way for a number of applications ranging for example from the development of single photon sensors in ultrathin film materials to highly efficient coupling of single photons to single plasmons for applications in quantum plasmonics.}\\

\newpage
\section{Methods}

\noindent \textbf{Experimental arrangement.} Single photon states are prepared by spontaneous parametric down conversion; a 100 mW laser diode centered at $\lambda$ = 405 nm (Cobolt MLD 405 nm laser diode module) is used to pump a 3 mm thick type-I beta barium borate (BBO) crystal producing non-collinear, degenerate entangled photon pairs via SPDC. The photon pairs are then coupled to single mode fibres by lenses and collimation objectives. A 10 nm bandpass filter centered at 810 nm is used to isolate the SPDC photons from any residual pump photons and ambient light. One of the fibres is connected directly to the photon counting apparatus (single photon avalanche detector (SPAD) and National Instruments counting card) and is used to herald the presence of another photon within the interferometer. The other fibre output is coupled to the input of a  interferometer. The polarisation state of the photons is set via a linear Glan-Taylor polarizer (extinction 10,000:1) prior to passing a lossless (50:50) non-polarising beam splitter. The beam splitter divides the single photons into arms $\alpha$ and $\beta$ of the interferometer, creating the quantum superposition state.\\

The counterpropagating photons from path $\alpha$ and $\beta$ are tightly focussed onto the metamaterial by $\times$10 Nikon microscope objectives producing a spot size of $\sim$5 $\mu$m. The sample is placed at the center of the interferometer, within the coherence length ($\sim$200 $\mu$m) of the photons and scanned via a piezo-electrically actuated linear translation stage over a few optical cycles. Following the sample, the photons are decoupled from the interferometer cavity via lossless 50:50 beamsplitters and coupled to multimode fibres. The photons are then detected in coincidence with the heralded photon via a second SPAD detector. The two detection arms ($\gamma$ and $\delta$), which are of different optical path length, are coupled into a single multimode fibre by means of a beamsplitter (not shown in Fig. \ref{fig2}) in order to measure the photon counts on a single SPAD detector. The output ports, $\gamma$ and $\delta$, are measured independently by blocking the beam before each of the fibre couplers or simultaneously with both channels open.\\

\begin{figure}[ht!]
\begin{centering}
\includegraphics[width = 8cm]{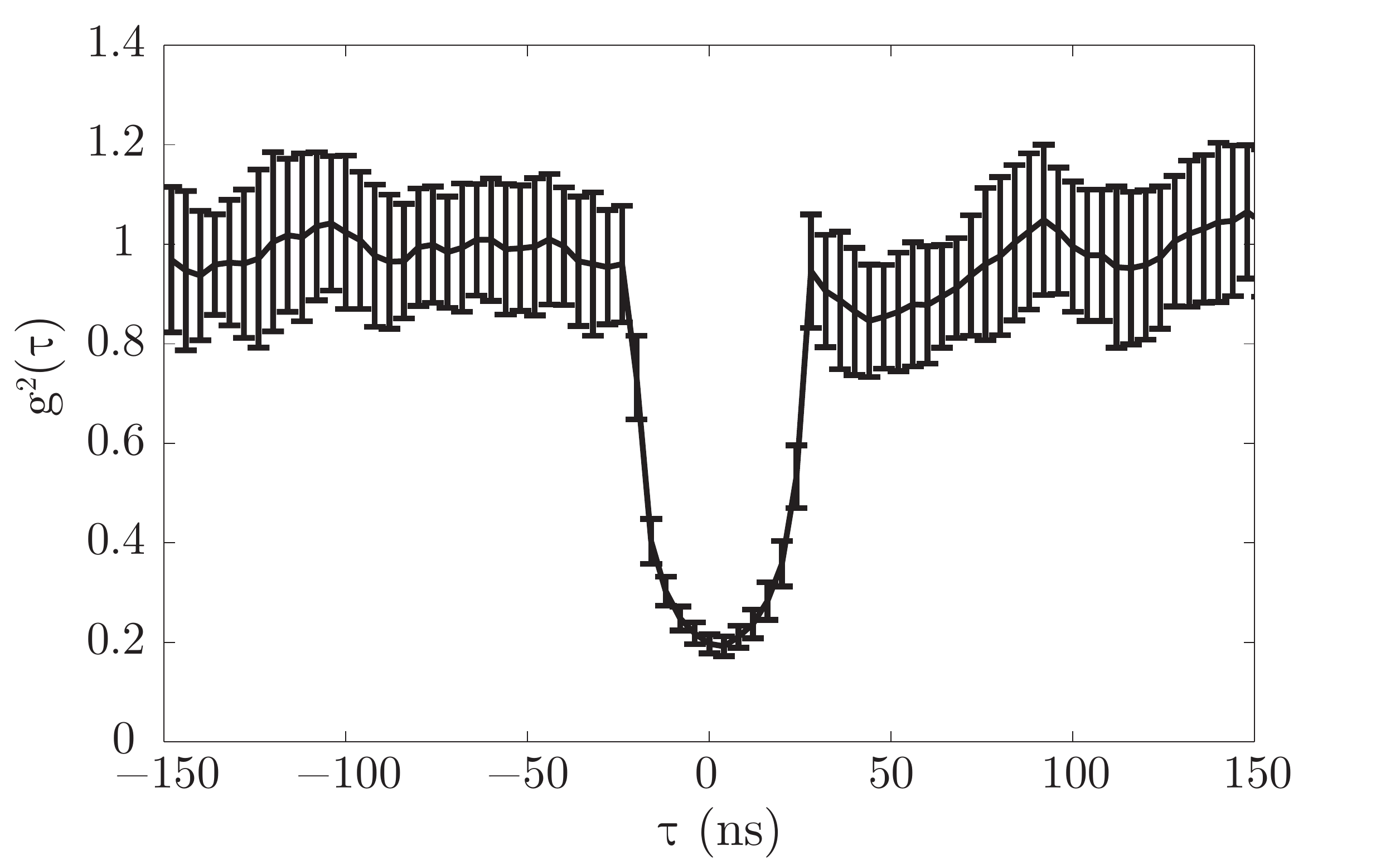}
\caption{\textbf{Experimental measurement of the single photon source $g^{(2)}$ function.} Measurements are taken with the same 25 ns integration window used in the coherent absorption experiments. The second order correlation function, $g^{(2)}$, is calculated as the mean of the total number of measurements taken at each time delay, $N = 30$. Error bars are calculated as the standard deviation over the total number of measurements taken.}
\label{fig_g2}
\end{centering}
\end{figure}

\noindent \textbf{Measurement of heralded single photon $g^{(2)}(\tau)$.}
Figure~\ref{fig_g2} shows the measured $g^{(2)}(\tau)$ for our heralded single-photon source. The standard $g^{(2)}(\tau)$ measurement uses three single-photon detectors: a gate detector $D_g$, which is used to herald the presence of a signal, and two detectors $D_1$ and $D_2$, one for each output of a beamsplitter placed in the path of the signal. A delay of time $\tau$ can be introduced to the path of detector $D_2$.  The resultant $g^{(2)}(\tau)$ is a function of the three-fold coincidence rate $R_{12g}(\tau)$, the two-fold coincidence rates $R_{1g}$ and $R_{2g}(\tau)$, and the single-event rate $R_g$; thus it is given by \cite{couteau}
\begin{equation}\label{g2}
g^{(2)}(\tau) =  \frac{R_{12g}(\tau)R_g}{R_{1g} R_{2g}(\tau)}.
\end{equation}
 As anticipated for a heralded single-photon source,  $g^{(2)}(0) < 0.5$ and $g^{(2)}(\tau \gg 0) =1$.  The width of dip is $\sim50$ ns, which is due to the 25 ns coincidence detection window and was chosen to be the same as that used in the coherent absorption experiments.  The data in Fig.~\ref{fig_g2} confirms the single-photon nature of our heralded source and thus the quantum nature of the perfect coherent absorption measurements. 

\noindent \textbf{Sample preparation.} The free-standing metamaterial structure with 50 $\mu$m x 50 $\mu$m overall area was fabricated in a 50 nm gold film by focused ion beam milling. Figure \ref{figs1} shows a SEM image of the sample and its unit cell structure and a comparison between absorption of the unstructured gold film and nanostructured metamaterial film of the same thickness.\\

\noindent \textbf{Definition of scattering matrix.} The matrix S of a infinitely thin absorbing layer has the general form:
\begin{eqnarray}
S=
\begin{pmatrix}
     t & r  & A    \\
     r & t  & B \\
     C & D & E
\end{pmatrix}
\end{eqnarray}
Conservation of energy imposes $|A|^2=|B|^2=1-|t|^2-|r|^2$. But we also have $2 Re(t r^{*})\pm |A|^2=0$, i.e. $2|t||r|{ \cos{\phi_{rt}}} = {\mp |A|^2}$, 
where $|A|^2$ represents the absorption of the film and $\phi_{rt}$ is the relative phase of $r$ and $t$.
Combining these equations, we obtain $t = 1 \pm r$ \cite{Thongrattanasiri}. 
Combining these relations fixes the maximum absorption for the travelling wave $|A|^2=0.5$, which is obtained for  
$t=\pm r$ and  (keeping only the negative sign term, i.e. assuming a $\pi$ phase shift in reflection) $t=0.5=-r$. 
 Finally the S matrix for a thin film perfect absorber takes the general form:
\begin{equation}
S = \left(\begin{array}{ccc}
0.5 & -0.5 & \sqrt{0.5}e^{i\rho}\\
-0.5 & 0.5 & \sqrt{0.5}e^{i\rho}\\
\sqrt{0.5}e^{i\sigma} & \sqrt{0.5}e^{i\sigma} & 0\end{array}\right)
\end{equation}
\noindent where $\rho$ and $\sigma$ are phases related to the excitation/de-excitation of the plasmon mode. Under the assumption that plasmon modes do not decay back into photon modes but are dissipated via non-radiating processes, these phases do not enter the measurable probabilities $P$.\\
It is also possible to demonstrate that the visibility of the total output energy only depends on the absorption $|A|^2$.
Using the equations above, the single channel visibility (symmetric thin film) is
$V=[{2 |t||r|}]/[{|t|^2+|r|^2}] =[{|A|^2}]/[{(|A|^2-1)\cos{\phi_{rt}}}]$. The total (i.e. summing over both output ports, $\gamma$ and $\delta$) output visibility is then given by:
\begin{equation}\label{Visibility}
V_{tot}=V \cos{\phi_{rt}}=\frac{|A|^2}{|A|^2-1}
\end{equation}
In order to obtain a visibility $>80\%$ as in our case, an absorption higher than $45 \%$ is required. \\
  We also explicitly verified numerically (commercial software, COMSOL) that for wavelengths with 50\% loss, the metamaterial induces an absolute phase shift on the reflected and transmitted beams of $\pi$ and $0$, respectively in agreement with the conditions $r=-t$ and $|r|= 1/2$ that must be satisfied with deeply subwavelength films and 50\% loss \cite{Thongrattanasiri}.

\section*{Acknowledgements}
This work is supported by the MOE Singapore (grant MOE2011-T3-1-005), the Leverhulme Trust, the Royal Society, the UK's Engineering and Physical Sciences Research Council through the Nanostructured Photonic Metamaterials Programme Grant and NTU-NAP startup grant No. M4080511. D. F. acknowledges financial support from the European Research Council under the European Unions Seventh Framework Programme (FP/2007-2013)/ERC GA 306559 and EPSRC (UK, Grant EP/J00443X/1). C. C. would like to thank the Champagne-Ardenne region for financial support via the visiting professor scheme and to thank the support of the French LABEX Action.
 S. V. acknowledges Y. Chong for useful discussions. The authors thank G. Adamo for helping with the fabrication and C. Altuzarra for helpful discussions. This work is part funded by the Ministry of Defence and is published with the permission of the Defence Science and Technology Laboratory on behalf of the Controller of HMSO.



\begin{thebibliography}{99}

\bibitem{Altewischer} E. Altewischer, M. P. van Exter, J. P. Woerdman, Plasmon-assisted transmission of entangled photons, Nature {\bf418}, 304-306 (2002).

\bibitem{Chang} D. E. Chang, A. S. S¿rensen, P. R. Hemmer, M. D. Lukin, Quantum optics with surface plasmons. Phys. Rev. Lett. {\bf97}, 053002 (2006).

\bibitem{Chang2} D. E. Chang, A.S. S¿rensen, E. A. Demler and M. D. Lukin, A single-photon transistor using nanoscale surface plasmons,  Nature Phys. {\bf3}, 807 (2007).

\bibitem{Akimov} A. V. Akimov, A. Mukherjee, C. L. Yu, D. E. Chang, A. S. Zibrov, P. R. Hemmer, H. Park, M. D. Lukin, Generation of single optical plasmons in metallic nanowires coupled to quantum dots. Nature. {\bf 450}, 402406 (2007).

\bibitem{Heeres} R. W. Heeres, L. P. Kouwenhoven, V. Zwiller, Quantum interference in plasmonic circuits. Nature Nanotechnology. {\bf8}, 719-722 (2013).

\bibitem{Tame} M.S. Tame, K. R. McEnery,  {\c{S}}. K. {\"O}zdemir, J. Lee, S. A. Maier, M. S. Kim, Quantum plasmonics. Nature Physics. {\bf 9}, 329 (2013).

\bibitem{Fakonas} J. S. Fakonas, H. Lee, Y. A. Kelaita and H. A. Atwater, Two-plasmon quantum interference, Nature Photonics {\bf 8}, 317 (2014).

\bibitem{Atwater} H. A. Atwater and A. Polman, Plasmonics for improved photovoltaic devices, Nature Mat.  {\bf 9}, 205 (2010).

\bibitem{Linic} S. Linic, P. Christopher and D. B. Ingram, Plasmonic-metal nanostructures for efficient conversion of solar to chemical energy, Nature Mat. {\bf 10}, 911 (2011).

\bibitem{Pincella} F. Pincella, K. Isozaki and K. Miki, A visible light-driven plasmonic photocatalyst, Light: Sci. $\&$ Appl. {\bf 3}, e133 (2014).

\bibitem{Wang} Y.-L. Wang, F. Nan, X.-L. Liu, L. Zhou, X.-N. Peng, Z.-K. Zhou, Y. Yu, Z.-H. Hao, Y. Wu, W. Zhang, Q.-Q. Wang, and Z. Zhang, Plasmon-Enhanced Light Harvesting of Chlorophylls on Near-Percolating Silver Films via One-Photon Anti-Stokes Upconversion, Sci. Rep. {\bf 3}, 1861 (2013).

\bibitem{Lee} M. Lee, J. U. Kim, J. S. Lee, B. I. Lee, J. Shin, and C. B. Park, Mussel-Inspired Plasmonic Nanohybrids for Light Harvesting, Adv. Mat. {\bf 26}, 4463 (2014).


\bibitem{Liu} X. Liu, T. Tyler, T. Starr, A. F. Starr, N. M. Jokerst, and W. J. Padilla, Taming the Blackbody with Infrared Metamaterials as Selective Thermal Emitters, Phys. Rev. Lett. {\bf 107}, 045901 (2011).

\bibitem{Kinkhabwala} A. Kinkhabwala, Z. Yu, Shanhui Fan, Y. Avlasevich, K. Mullen and W. E. Moerner , Large single-molecule fluorescence enhancements produced by a bowtie nanoantenna, Nature Phot. {\bf 3}, 654 (2009).

\bibitem{Fedetov} V. A. Fedotov, P. L. Mladyonov, S. L. Prosvirnin, and N. I. Zheludev, Planar electromagnetic metamaterial with a fish scale structure, Phys. Rev. E {\bf 72}, 056613 (2005).

\bibitem{Schwanecke} A. S. Schwanecke, V. A. Fedotov, V. V. Khardikov, S. L. Prosvirnin, Y. Chen, and N. I. Zheludev, Optical magnetic mirrors, J. Opt. A {\bf 9}, L1-L2 (2007).

\bibitem{Liu2} N. Liu, L. Langguth, T. Weiss, J. KŠstel, M. Fleischhauer, T. Pfau, and H. Giessen, Plasmonic analogue of electromagnetically induced transparency at the Drude damping limit, Nature Mat. {\bf8}, 758 (2009).

\bibitem{Thongrattanasiri} S. Thongrattanasiri, F. H. L. Koppens, F. J. Garcia de Abajo, Complete optical absorption in periodically patterned graphene, Phys. Rev. Lett. {\bf108}, 047401 (2012).

\bibitem{Chong} Y. D. Chong, Li Ge, Hui Cao, and A. D. Stone, Coherent Perfect Absorbers: Time-Reversed Lasers, Phys. Rev. Lett. {\bf105}, 053901 (2010).

\bibitem{Wan} W. Wan, Y. D. Chong, L. Ge, H. Noh, A. D. Stone, H. Cao, Time-Reversed Lasing and Interferometric Control of Absorption, Science {\bf331}, 889 (2011).

\bibitem{Dutta-Gupta} S. Dutta-Gupta, O. J. F. Martin, S. Dutta Gupta, and G. S. Agarwal, Controllable coherent perfect absorption in a composite film, Opt. Exp. {\bf20}, 1330 (2012).

\bibitem{Zheludev} J. Zhang, K.F. MacDonald, N. I Zheludev, Controlling light-with-light without nonlinearity. Light: Sci. \& Appl. {\bf 1}, e18 (2012)

\bibitem{Fang} X. Fang, M. L. Tseng, J.-Y. Ou, K. F. MacDonald, D. P. Tsai, N. I. Zheludev, Ultrafast all-optical switching via coherent modulation of metamaterial absorption. Appl. Phys. Lett. {\bf104}, 141102 (2014)

\bibitem{Mousavi} S. A. Mousavi, E. Plum, J. Shi, N. I. Zheludev, Coherent control of optical activity and optical anisotropy of thin metamaterials. Appl. Phys. Lett. {\bf105}, 011906 (2014).


\bibitem{Jeffers} J. Jeffers, Interference and the lossless lossy beam splitter. Journal of Modern Optics. {\bf47}, 11, 1819-1824 (2000)

\bibitem{Khardikov}  V. V. Khardikov, E. O. Iarko, S. L. Prosvirnin, Trapping of light by metal arrays. J. Opt. {\bf 12}, 045102 (2010)

\bibitem{Davis} T. J. Davis, D. E. G\'{o}mez, and F. Eftekhari, All-optical modulation and switching by a metamaterial of plasmonic circuits, Optics Letters, Vol. {\bf39}, Issue 16, pp. 4938-4941 (2014)


\bibitem{couteau} E. Bocquillon, C. Couteau, M. Razavi, R. Laflamme, and G. Weihs, Coherence measures for heralded single-photon sources, Phys. Rev. A {\bf 79}, 035801 (2009)



\end{thebibliography}
\end{document}